# Distinguishing Among Strong Decay Models

Paul Geiger

*Department of Physics, Carnegie Mellon University, Pittsburgh, PA 15213.*

Eric S. Swanson

*Center for Theoretical Physics, Laboratory of Nuclear Science, and Department of Physics,*

*Massachusetts Institute of Technology, Cambridge, MA 02139.*

(May, 1994)

## Abstract

Two competing models for strong hadronic decays, the $^3P_0$ and $^3S_1$ models, are currently in use. Attempts to rule out one or the other have been hindered by a poor understanding of final state interactions and by ambiguities in the treatment of relativistic effects. In this article we study meson decays in both models, focussing on certain amplitude ratios for which the relativistic uncertainties largely cancel out (notably the $S/D$ ratios in $b_1 \to \pi\omega$ and $a_1 \to \pi\rho$), and using a Quark Born Formalism to estimate the final state interactions. We find that the $^3P_0$ model is strongly favoured. In addition, we predict a $P/F$ amplitude ratio of $1.6 \pm .2$ for the decay $\pi_2 \to \pi\rho$. We also study the parameter-dependence of some individual amplitudes (as opposed to amplitude ratios), in an attempt to identify a "best" version of the $^3P_0$ model.

# I. INTRODUCTION

Quark pair-creation models for the strong decays of hadrons [1] have have been formulated and studied by a number of authors, beginning in about 1969. The Orsay group [2] developed Micu's original suggestion [3] that strong decays proceed by simple quark rearrangement following the creation of a $^3P_0$ $q\bar{q}$ pair from the vacuum. They applied this model with considerable success to a number of baryon and meson decays. The most extensive application to meson decays was made by Kokoski and Isgur in 1987 (Ref. [4], hereafter KI). In addition to calculating over 400 different amplitudes (about 60 of which have so far been measured) these authors also placed the model on a firmer theoretical footing by showing how it could be derived from a flux-tube picture based on lattice QCD. Strong baryon decays were also studied in the $^3P_0$ model by Stancu and Stassart [5] and by Capstick and Roberts [6].

Intriguingly, a quite different pair-creation model, also rooted in a flux tube picture of confinement, was developed concurrently with the $^3P_0$ model. It is based on the observation that the QCD interaction Hamiltonian, $H_{\text{int}} \sim \int \bar{\psi}\gamma_0 \, \boldsymbol{\lambda} \cdot \mathbf{A}_0 \psi$, when applied to an oriented chromoelectric flux tube, breaks the flux tube and creates a $^3S_1$ $q\bar{q}$ pair. Alcock $et\ al.$ [7] have shown that this $^3S_1$ model provides a good description of $J/\Psi$ and $\Upsilon$ decay widths (as does the $^3P_0$ model). Kumano [8] studied pion-nucleon couplings in both models, and found the data unable to discriminate between them. A similarly inconclusive situation exists in $p\bar{p}$ scattering, where the $^3S_1$ and $^3P_0$ models appear to describe $p\bar{p} \to \Lambda\bar{\Lambda}$ equally well [9].

It was pointed out in KI that S-wave decays of mesons provide a good arena for contrasting the two models, since the predicted rates for such decays tend to be much larger in the $^3S_1$ model. In fact, KI concluded that the measured $S/D$ amplitude ratio in $b_1 \to \pi\omega$ rules out a significant $^3S_1$ component in the decay operator: the $^3S_1$ model predicts $S/D \sim 20$, whereas the measured ratio (and the $^3P_0$ prediction) is about 4. (Note that it is almost impossible to salvage the $^3S_1$ prediction by tuning the model, since amplitude *ratios* tend to be quite parameter-independent.)

However, Kumano and Pandharipande [10] (hereafter KP), argued for a reprieve of the $^3S_1$ model, because final state interactions (FSI's) among the decay products (which had not been considered in any of the previous studies) can substantially alter the amplitude predictions. They showed that, for example, a repulsive nonresonant background FSI in $\pi\omega$ (which they modelled by a hard-core potential with a core radius of about 0.4 fm) significantly reduces S-wave amplitudes and can lower the above $S/D$ prediction to the experimental value. Because of the dearth of experimental and theoretical information on short-range meson interactions, KP were unable to motivate their FSI potentials (which were tuned to fit the measured decay amplitudes) and could only argue that they were similar to empirical nucleon-nucleon potentials, and hence not unreasonable.

In this paper we propose to employ the Quark Born Formalism introduced in Ref. [11,12] to *calculate* meson-meson FSI's in various channels, and then to re-examine the KP suppression mechanism using these calculated FSI's. The Quark Born Formalism has been successfully applied to $\pi\pi$ [11,12], $K\pi$ [13], $KN$ [14] and $NN$ [15] scattering so that we expect our FSI estimates to be reasonable. We are able to take into account some effects not considered in KP, such as the difference in FSI's among different $\pi\pi$ isospin channels. We shall find that our FSI potentials have a much smaller effect on decay amplitudes than the rather drastic hard-core potentials used in KP. Moreover, many of our potentials are *attractive* rather than repulsive. By focussing on amplitude ratios such as the above- mentioned $S/D$ ratio in $b_1 \to \pi\omega$, as well as the corresponding ratio in $a_1 \to \pi\rho$ (recently measured by the Argus collaboration [16]), we shall be able to sidestep many of the uncertainties associated with normalization and relativistic effects. This will enable us to draw fairly firm conclusions regarding the validity of each model.

The paper is organized as follows: In Section II we provide descriptions of the decay models and of a brief review of the Quark Born Formalism. Section III contains our main results: effective potentials describing the mesonic FSI's, amplitude ratios as a function of FSI strengths, and a study of the parameter-dependence of some individual decay amplitudes. We summarize and conclude in Section IV.

## II. MODELS AND METHODS

A flux-tube-breaking picture of meson decays underlies both the $^3P_0$ and $^3S_1$ models. In turn, the flux-tube picture is suggested by (the strong coupling limit of) Hamiltonian lattice QCD. (Refs. [4,17] discuss these matters in detail; we shall only provide a brief recap here.) The lattice QCD Hamiltonian contains a flux-tube-breaking term which creates a quark and antiquark at neigbouring lattice sites $\mathbf{R}$ and $\mathbf{R} + a\hat{\mathbf{n}}$ with the effective operator

$$C(\mathbf{R}, \hat{\mathbf{n}}) = \Psi^\dagger(\mathbf{R})\, \boldsymbol{\alpha} \cdot \hat{\mathbf{n}}\, \Psi(\mathbf{R} + a\hat{\mathbf{n}}) \tag{1}$$

(where $a$ is the lattice spacing). For the medium-to-small lattice spacings relevant to hadron spectroscopy, this may be expanded as

$$C(\mathbf{R}, \hat{\mathbf{n}}) \approx \Psi^\dagger(\mathbf{R})\, \boldsymbol{\alpha} \cdot \hat{\mathbf{n}}\, \Psi(\mathbf{R}) + a\, \Psi^\dagger(\mathbf{R})\, \boldsymbol{\alpha} \cdot \hat{\mathbf{n}}\, \hat{\mathbf{n}} \cdot \boldsymbol{\nabla} \Psi(\mathbf{R}) \tag{2}$$

If the flux tube is "rough" at the scale $a$ (*i.e.*, if its zero-point oscillations are so strong that the flux tube meanders, with essentially random orientation at each point), then one should average over $\hat{\mathbf{n}}$. Only the second term in (2) survives such averaging, so that the effective pair creation operator becomes

$$H^P(\mathbf{R}) \equiv \gamma_P\, \Psi^\dagger(\mathbf{R})\, \boldsymbol{\alpha} \cdot \overleftrightarrow{\boldsymbol{\nabla}}\, \Psi(\mathbf{R})\,, \tag{3}$$

and the $q\bar{q}$ pairs are created with $^3P_0$ quantum numbers. The parameter $\gamma_P$ introduced in this expression is a phenomenological constant representing the intrinsic pair creation strength. It will be adjusted to fit decay data. (Similar comments apply to $\gamma_S$ in Eq. (4) below.)

If, on the other hand, the flux tube is essentially straight then the first term in (2) survives and leads to pair creation in a $^3S_1$ state:

$$H^S(\mathbf{R}, \hat{\mathbf{n}}) \equiv \gamma_S\, \Psi^\dagger(\mathbf{R})\, \boldsymbol{\alpha} \cdot \hat{\mathbf{n}}\, \Psi(\mathbf{R})\,. \tag{4}$$

We call the above operators "effective" because they are built from quark fields only, with no explicit reference to the flux tubes that they break. Consider applying these operators to a "frozen" meson state $|\mathbf{r}_1, \mathbf{r}_2\rangle$. (This ket is meant to represent a quark at $\mathbf{r}_1$ and an antiquark at $\mathbf{r}_2$, joined by a ground-state flux tube. Spin indices are suppressed for the moment). We have

$$H^P(\mathbf{R})|\mathbf{r}_1,\mathbf{r}_2\rangle = \gamma_P F^P(\mathbf{r},\mathbf{w})\Psi^\dagger(\mathbf{R})\,\boldsymbol{\alpha}\cdot\overleftrightarrow{\nabla}\,\Psi(\mathbf{R})|\mathbf{r}_1,\mathbf{r}_2\rangle \qquad (5)$$

$$H^S(\mathbf{R},\hat{\mathbf{n}})|\mathbf{r}_1,\mathbf{r}_2\rangle = \gamma_S F^S(\mathbf{r},\mathbf{w})\Psi^\dagger(\mathbf{R})\,\boldsymbol{\alpha}\cdot\hat{\mathbf{n}}\,\Psi(\mathbf{R})|\mathbf{r}_1,\mathbf{r}_2\rangle\,, \qquad (6)$$

with $\hat{\mathbf{n}}$ parallel to $\mathbf{r}_1 - \mathbf{r}_2$; the coordinates are defined in Fig. 1. The flux-tube information is contained in the $F$'s, which give the overlap of an initial, unbroken flux tube with two final-state flux tubes. In the $^3S_1$ model, because the flux tube is assumed to be essentially straight, $F^S$ is usually taken to vanish unless the pair creation point $\mathbf{R}$ lies along the line joining $\mathbf{r}_1$ and $\mathbf{r}_2$. In the $^3P_0$ model, the oscillating flux tube implies a more complicated form for $F^P$. Based on their analysis of an idealized (harmonic) flux tube, KI advocates the approximation $F^P = e^{-(b/2)w_{\min}^2}$, where $w_{\min}$ is the shortest distance from the pair creation point to the line joining $Q$ and $\bar{Q}$ in Fig. 1, and $b$ is the string tension in the flux tube. Typically, one uses the spectroscopic value $b \approx 0.18$ GeV$^2$, but it turns out that calculated decay widths are not very sensitive to this parameter. In fact, in the original $^3P_0$ model $F^P$ was implicitly set equal to one [2]. For the sake of completeness, our calculations in both the $^3P_0$ and $^3S_1$ models were carried out with three kinds of flux tube:

$$\begin{aligned}\text{narrow flux tube}: \quad & F = \delta(w_{\min}) \\ \text{medium flux tube}: \quad & F = e^{-(b/2)w_{\min}^2} \\ \text{wide flux tube}: \quad & F = 1.\end{aligned} \qquad (7)$$

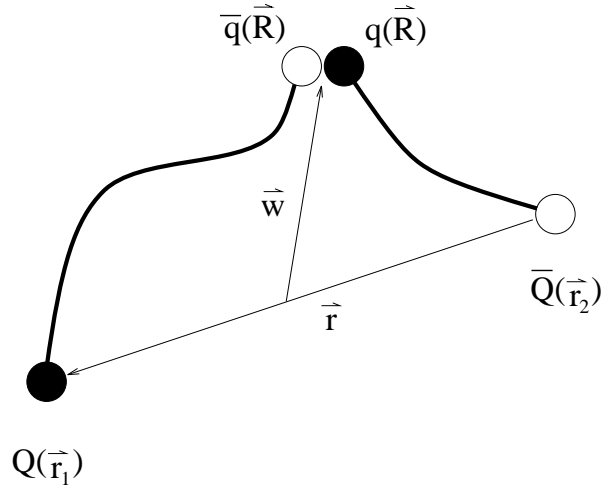

Fig. 1. Coordinates for meson decay by flux-tube breaking.

The non-relativistic limits of $H^P$ and $H^S$ are (keeping only creation operators)

$$H^P(\mathbf{R}) \to \gamma_P \sum_{s_3,s_4} \vec{\chi}_{s_3 s_4}\, b^\dagger_{s_3}(\mathbf{R})\, \overleftrightarrow{\nabla}\, d^\dagger_{s_4}(\mathbf{R}) \qquad (8)$$

$$H^S(\mathbf{R},\hat{\mathbf{n}}) \to \gamma_S \sum_{s_3,s_4} \vec{\chi}_{s_3 s_4}\, b^\dagger_{s_3}(\mathbf{R})\, \hat{\mathbf{n}}\, d^\dagger_{s_4}(\mathbf{R}) \qquad (9)$$

where

$$\vec{\chi}_{s_3 s_4} \equiv \begin{pmatrix} 2\delta_{s_3\uparrow}\delta_{s_4\uparrow} \\ \\ -\delta_{s_3\downarrow}\delta_{s_4\uparrow} - \delta_{s_3\uparrow}\delta_{s_4\downarrow} \\ \\ -2\delta_{s_3\downarrow}\delta_{s_4\downarrow} \end{pmatrix}. \qquad (10)$$

Decay amplitudes are given by $\langle BC|H^{(P\ or\ S)}|A\rangle$, where $|A\rangle$, $|B\rangle$, $|C\rangle$ are quark-model meson states:

$$|A\rangle = \sum_{\substack{s_1 s_2 \\ f_1 f_2}} \chi^A_{s_1 s_2} \Phi^A_{f_1 f_2} \int d^3 r_1 d^3 r_2\, \frac{1}{(2\pi)^{3/2}} e^{i\mathbf{P}_A\cdot(\mathbf{r}_1+\mathbf{r}_2)/2}\, \psi_A(\mathbf{r}_1-\mathbf{r}_2)\, |\mathbf{r}_1 \mathbf{s}_1 f_1;\, \mathbf{r}_2 \mathbf{s}_2 f_2\rangle,\ \ etc. \quad (11)$$

Colour indices have been suppressed. Our normalization is $\langle A(\mathbf{P}')|A(\mathbf{P})\rangle = \delta^3(\mathbf{P}-\mathbf{P}')$.

The decay amplitude may be written as

$$\langle BC|H^{(P\ or\ S)}|A\rangle = \gamma_{(P\ or\ S)}\, \phi \vec{\Sigma}\cdot\vec{I}, \qquad (12)$$

where $\phi$ is a flavour overlap,

$$\phi \equiv \sum_{f_1 f_2 f_3 f_4} \Phi^{*B}_{f_3 f_2} \Phi^{*C}_{f_1 f_4} \Phi^{A}_{f_1 f_2} , \qquad (13)$$

$\vec{\Sigma}$ is a spin overlap which can be expressed in terms of the meson spin wavefunctions and $\vec{\chi}$ as

$$\vec{\Sigma} \equiv \sum_{s_1 s_2 s_3 s_4} \chi^{*B}_{s_3 s_2} \chi^{*C}_{s_1 s_4} \chi^{A}_{s_1 s_2} \vec{\chi}_{s_3 s_4} , \qquad (14)$$

and $\vec{I}$ is a spatial overlap:

$$\vec{I}^P = \int d^3r d^3w \; \psi_A(\mathbf{r}) \; F^P(\mathbf{r}, \mathbf{w}) \left( \vec{\nabla}_B + \vec{\nabla}_C + \vec{\nabla}_{BC} \right) \psi^*_B(\frac{\mathbf{r}}{2} + \mathbf{w}) \; \psi^*_C(\frac{\mathbf{r}}{2} - \mathbf{w}) \; \psi^*_{BC}(-\frac{\mathbf{r}}{2}) \quad (15)$$

$$\vec{I}^S = \int d^3r d^3w \; \psi_A(\mathbf{r}) \; F^S(\mathbf{r}, \mathbf{w}) \; \hat{\mathbf{r}} \; \psi^*_B(\frac{\mathbf{r}}{2} + \mathbf{w}) \; \psi^*_C(\frac{\mathbf{r}}{2} - \mathbf{w}) \; \psi^*_{BC}(-\frac{\mathbf{r}}{2}) \qquad (16)$$

for the $^3P_0$ and $^3S_1$ models, respectively. Equations (15) and (16) are valid when all the quarks participating in a decay have the same mass, as is the case for the processes we consider here. (In deriving these expressions, our convention is that meson $B$ is the one which contains the *antiquark* from meson $A$. Of course the other diagram must be included in all calculations.)

We calculate decay amplitudes using two distinct sets of meson wavefunctions:

(i) Harmonic oscillator wavefunctions with a single oscillator parameter, $\beta = 0.4$ GeV, for all mesons. Although these cannot be regarded as realistic, they are a *de facto* standard for many nonrelativistic quark model calculations, and they lead to analytic results for many of our calculations, thus providing a check on our computer code. In addition, the quite good results obtained with SHO wavefunctions indicate that our findings are not strongly tied to a particular choice of wavefunctions.

(ii) More realistic wavefunctions, obtained as eigenstates of a nonrelativistic Hamiltonian that incorporates Coulomb and linear-confinement terms, and a "smeared" magnetic hyperfine term. This Hamiltonian and its parameters (which were fitted to the low-lying meson spectrum) are described fully in Ref. [12].

The final ingredient in equations (15) and (16) is $\psi_{BC}$, the relative wavefunction of the decay products. In the absence of final state interactions $\psi_{BC}$ is essentially just a Bessel function:

$$\psi_{BC}(\mathbf{r}) = \sqrt{\frac{2}{\pi}}\, j_\ell(kr)\, Y_{\ell m}(\Omega_r) \qquad (17)$$

(where $k$ is the decay momentum). When FSI's are turned on, $j_\ell$ is replaced by $u_\ell$, the solution of a radial Schrödinger equation in the presence of the effective FSI potential $V_{\text{eff}}$. This potential simulates the nonresonant "background" contribution to the interaction of the outgoing mesons. We calculate it using the Quark Born Formalism [11,12], which involves computing the lowest order quark-level T-matrix for a given scattering process and equating it to an effective T-matrix for pointlike mesons. At lowest order, a single t-channel gluon is exchanged between quarks which have been placed in mesonic bound state wavefunctions. Colour neutrality then necessitates the exchange of a pair of quarks, hence the effective potentials are of short range; we neglect long range meson-exchange contributions. Nonlocality is incorporated by allowing for different potentials in different partial waves.

The method has proven surprisingly effective in channels where meson exchange and resonance production do not interfere with the quark-level interaction [11–15]. The method predicts a repulsive interaction for pions in the isotensor channel and an attractive interaction in the isosinglet channel, in agreement with $\pi\pi$ scattering data (and in contrast with the use of isospin-independent $\pi\pi$ core interactions in Ref. [10]).

For the final states we shall be considering here, the effective potential also receives contributions from diagrams involving $q\bar{q}$ annihilation. The Quark Born Formalism has recently been extended to include such contributions [18]; the quark-level interaction is given by the nonrelativistic limit of s-channel one gluon exchange:

$$H_{\text{ann}} = f \cdot \frac{2\pi\alpha_s}{m^2}\left(\frac{3}{4} + \vec{S}_i \cdot \vec{S}_j\right) \delta(\vec{r}_{ij})\, (\lambda_i^a/2) \cdot (\lambda_j^a/2). \qquad (18)$$

This operator connects the incoming 2-meson state to a $q\bar{q}g$ (hybrid) intermediate state. As discussed in Ref. [18], a perturbative description of this intermediate state is inadequate;

colour confinement must at least raise the $q\bar{q}g$ mass from its perturbative value of $2m_q$ to the physical hybrid mass. One simple way of modelling such presently uncalculable confinement effects is to introduce an effective gluon mass into the gluon propagator:

$$s^{-1} \to \left(s - \mu_g^2\right)^{-1} . \tag{19}$$

The parameter $f$ in Eq. (18) is intended to incorporate this modification. It is fixed by comparison to $I = 0$ $\pi\pi$ scattering.

An alternative, probably better, way to proceed would be to calculate the meson-meson-hybrid couplings in a more realistic model such as the flux tube model of Ref. [17]. However, such a calculation appears to be extremely difficult since it involves summing over many excited hybrid intermediate states (S-wave mesons decouple from the lowest lying hybrids, and the excited hybrid spectrum is quite dense [17,19].) One may expect that the rather high hybrid masses will yield a small annihilation coupling; however, for the purposes of this paper we seek an upper bound on the coupling. Thus we adopt the first method instead. Its successful application to $\pi\pi$ scattering leads us to expect, on the basis of SU(6) symmetry, that it will provide a reasonable description of the ($\pi\omega$ and $\pi\rho$) channels we shall need to consider.

Finally, an $A \to BC$ decay width is given by

$$\Gamma = \gamma_{(P \text{ or } S)} \, 2\pi k \, (\text{Ph.Sp.}) \int d\Omega_q |\mathcal{M}(\mathbf{k})|^2 \tag{20}$$

where $\mathcal{M}$ is defined by

$$\delta^3(\mathbf{P}_A) \, \mathcal{M}(\mathbf{k}) = \langle B(\mathbf{k}) \, C(-\mathbf{k}) | H^{(P \text{ or } S)} | A(\mathbf{P}_A) \rangle. \tag{21}$$

As discussed in Refs. [4,10], the choice of phase space (Ph.Sp. in equation (20)) is not entirely clear. The decay models we use are nonrelativistic, so it is perhaps most consistent to use nonrelativistic phase space, Ph.Sp. $= M_B M_C / M_A$. However in many cases the decay momenta are quite large so that the actual phase space, Ph.Sp. $= E_B E_C / M_A$, is significantly different from the nonrelativistic limit. The authors of Ref. [4] employ a third option,

Ph.Sp. = $\tilde{M}_B \tilde{M}_C / \tilde{M}_A$, where $\tilde{M}_i$ is the calculated mass of meson $i$ in a spin-independent quark model. (This form is intended to interpolate between the weak binding limit where the model is exact, and the quite strongly bound meson states that one must actually use in equation (20) ). Following the approach of Ref. [10], we perform our calculations for all three types of phase space; the differences in the results may be taken to indicate the inherent uncertainty of the models. (Note that the $S/D$ amplitude ratios are of course completely independent of the choice of phase space.)

## III. RESULTS AND DISCUSSION

Like the decay amplitudes, the final state interactions are obtained as overlaps of meson wavefunctions. Thus we have two cases to consider, corresponding to the SHO and Coulomb-plus-linear wavefunctions described above.

The T-matrices in the SHO case are as follows [20]. For $\pi\omega$ scattering

$$T_{\text{ex}}^{\text{hyp}} = \frac{4\pi\alpha_s}{27m^2} \left[ 3e^{-x(1-\mu)} - e^{-x(1+\mu)} + \frac{16}{3\sqrt{3}} e^{-4x/3} \right] \quad (22)$$

$$T_{\text{ex}}^{\text{quad}} = \frac{\pi^{3/2}}{3\sqrt{2}} \frac{1}{\beta m} \left( 3 - 4x \right) e^{-2x} \quad (23)$$

$$T_{\text{ann}} = f \cdot \frac{2\pi\alpha_s}{9m^2} \left[ 3e^{-x(1-\mu)} - e^{-x(1+\mu)} \right] \quad (24)$$

where $m$ is the quark mass, $x = k^2/4\beta^2$, $\mu$ is the cosine of the centre-of-mass scattering angle, and $f$ is the parameter discussed in Section II. $T_{\text{ex}}^{\text{hyp}}$ and $T_{\text{ex}}^{\text{quad}}$ are the hyperfine and quadratic confinement contributions to the $t$-channel T-matrix, respectively, while $T_{\text{ann}}$ is the $s$-channel T-matrix. It is interesting to note that the confinement term is comparable in strength to the hyperfine term; in most cases studied previously (such as $NN$ scattering), the hyperfine term is dominant (see Ref. [12] for further comments on this).

For isospin-1 $\pi\rho$ scattering the results are

$$T_{\text{ex}}^{\text{hyp}} = T_{\text{ex}}^{\text{conf}} = 0 \tag{25}$$

$$T_{\text{ann}} = f \cdot \frac{4\pi\alpha_s}{9m^2} \left[ 3e^{-x(1-\mu)} + e^{-x(1+\mu)} \right] , \tag{26}$$

and for $\pi\pi$ scattering in the $\rho$ channel,

$$T_{\text{ex}}^{\text{hyp}} = T_{\text{ex}}^{\text{conf}} = 0 \tag{27}$$

$$T_{\text{ann}} = f \cdot \frac{4\pi\alpha_S}{3m^2} \left[ e^{-x(1-\mu)} - e^{-x(1+\mu)} \right] . \tag{28}$$

These T-matrix elements are to be simulated by effective potentials of the form $V_{\text{eff}} = a \exp[-1/2(r/b)^2]$. The Gaussian parameters $a$ and $b$ are obtained by equating the low-momentum behaviour of the actual and effective T-matrices, partial-wave by partial-wave [21]. For $\pi\omega$,

$$a_{\text{ex}}^{\text{hyp}}(\ell = 0) = \frac{\sqrt{2}\alpha_s}{27\sqrt{\pi}m^2 b^3}(2 + \frac{16}{3\sqrt{3}}) \tag{29}$$

$$a_{\text{ex}}^{\text{quad}}(\ell) = \frac{1}{4\beta m b^3}\delta_{\ell,0} \tag{30}$$

$$a_{\text{ann}}(\ell) = \frac{f\alpha_s}{\sqrt{2\pi} 9 m^2 b^3}(3 - (-)^\ell). \tag{31}$$

For $\pi\rho$ one obtains

$$a_{\text{ann}}(\ell) = \frac{2f\alpha_s}{\sqrt{2\pi} 9 m^2 b^3}(3 + (-)^\ell). \tag{32}$$

And for $\pi\pi$

$$a_{\text{ann}}(\ell = \text{odd}) = \frac{\sqrt{2}f\alpha_S}{\sqrt{\pi} 3 m^2 b^3} . \tag{33}$$

For the S-wave $\pi\omega$ hyperfine potential, $b \approx 1.1/(2\beta)$; for the $\omega\pi$ confinement potential $b \approx 0.9/\beta$; in all other cases $b = 1/(2\beta)$. Table I shows the results of numerically evaluating these expressions with typical quark model parameters: $\beta = 0.4$ GeV (as in Ref. [4]), $m = 0.33$ GeV, and $\alpha_s = 0.6$, and $f = -2.573$ (from Ref. [18]).

Also shown in Table I are the various T-matrices evaluated with Coulomb-plus-linear wavefunctions. (The results for $T_{\text{ex}}$ were previously reported in Ref. [12].) The quark potential parameters in this case were determined by a fit to meson masses. Although there is some variation between the SHO and C+L columns, these will have only minor effects on the decay amplitudes. Observe that the confinement effective potentials are very similar in the SHO and C+L cases. This has also been noted in Ref. [12] in the case of $\rho\rho$ scattering; it seems that the specific form of the confinement potential is unimportant – only the fact that it confines with a strength determined by the meson spectrum matters.

TABLE I. Effective Potential Parameters.

|  | $a_{\text{SHO}}$ (GeV) | $b_{\text{SHO}}$ (GeV$^{-1}$) | $a_{\text{C+L}}$ (GeV) | $b_{\text{C+L}}$ (GeV$^{-1}$) |
|---|---|---|---|---|
| $V_{\text{ex,hyp}}^{\omega\pi}$ | 0.32 | 1.37 | 0.29 | 1.55 |
| $V_{\text{ex,conf}}^{\omega\pi}$ | 0.16 | 2.28 | 0.11 | 2.07 |
| $V_{\text{ann}}^{\omega\pi}(\ell = \text{even})$ | -0.64 | 1.25 | -0.56 | 1.35 |
| $V_{\text{ann}}^{\rho\pi}(I = 1, \ell \text{ even})$ | -2.56 | 1.25 | -2.24 | 1.35 |
| $V_{\text{ann}}^{\pi\pi}(I = 1, \ell \text{ odd})$ | -1.93 | 1.25 | -2.55 | 1.11 |

We now turn to the crucial $S/D$ ratios in $b_1 \to \pi\omega$ and $a_1 \to \pi\rho$. We isolate the dependence of these quantities on the FSI strengths by fixing the Gaussian width parameters at $b = 1.5$ GeV$^{-1}$. Also, the D-wave effective potentials are set equal to the corresponding S-wave ones so that we may graph $S/D$ versus a single independent variable. (Almost any other *Ansatz* for the $D$-wave potentials could be used instead; the distortion induced in higher waves by any reasonable $V_{\text{eff}}$ is small). Figure 2 shows the $S/D$ amplitude ratio for $b_1 \to \pi\omega$ versus the FSI strength $a$, for various wavefunction and flux-tube choices. The arrow indicates our "canonical" value of $a$, based on the Table I results. The experimental amplitude ratio is represented by a shaded horizontal band [22]. Figure 3 shows the corresponding results for $a_1 \to \pi\rho$ [23].

The figures show that extremely large FSI's are required to bring the $^3S_1$ ratios down to the experimental bands. On the other hand, the $^3P_0$ results are close to the data over a range of small to medium FSI strengths on the order of our calculated strengths. The $^3P_0$ ratios are not very sensitive to the assumed form of the wavefunctions, however the SHO wavefunctions prefer a medium to wide flux tube while the Coulomb-plus-linear wavefunctions require a medium to narrow flux tube. Notice that the $^3P_0$ ratios are much more sensitive to the flux tube width than are the $^3S_1$ ratios. Our results for the $P/F$ ratio in $\pi_2 \to \pi\rho$ show similar variation with respect to wavefunctions and flux tube geometry. In the region of parameter space that fits the measured ratios best, $P/F \approx 1.6$. We (subjectively) estimate the error on this prediction to be $\pm 0.2$

The ratio $[\Gamma(\pi_2 \to \rho\pi)/\Gamma(\rho_3 \to \omega\pi)]^{1/2}$, shown in Fig. 4, is only slightly more model-dependent. Because of the near degeneracy of the masses in these two decays ($m_{\rho_3} = 1691 \pm 5$ and $m_{\pi_2} = 1670 \pm 20$ [22]), the three phase space *Ansätze* described earlier give results that are equal to within 1%. The $\rho_3$ decay is an F-wave decay, consequently it is almost completely insensitive to FSI's. We therefore set $V_{\text{eff}}^{\omega\pi} = 0$ and graph the ratio as a function of the $\rho\pi$ potential. (The $\pi_2 \to (\rho\pi)$ has $P$ and $F$-wave components; only the former are sensitive to $V_{\text{eff}}^{\rho\pi}$.)

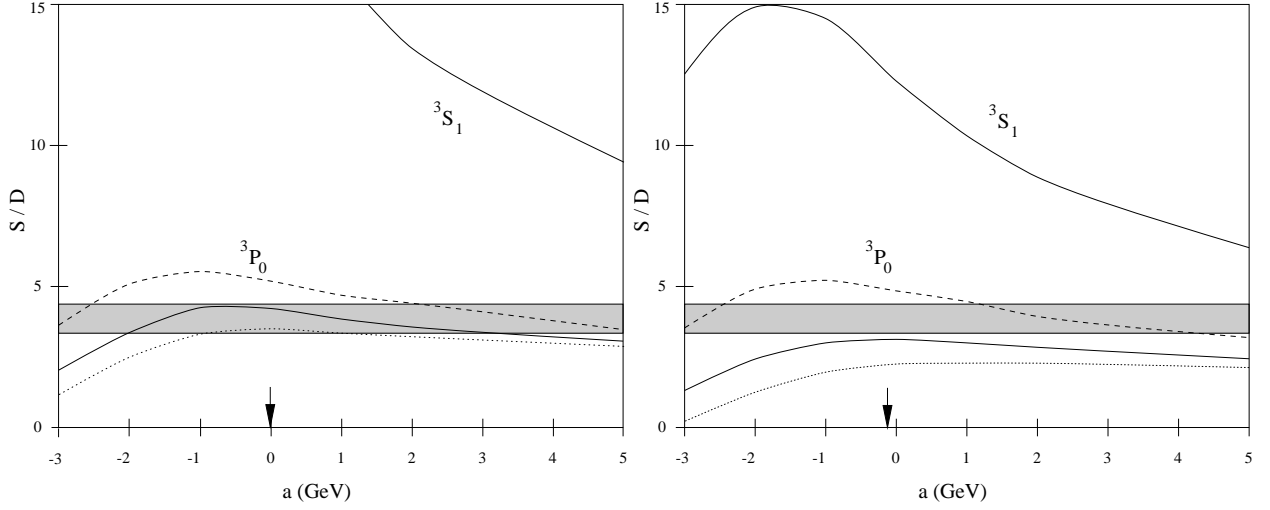

Fig. 2. $S/D$ amplitude ratios for $b_1 \to \pi\omega$. In the left graph, SHO wavefunctions were used; in the right, Coulomb-plus-linear wavefunctions. Dashed lines show narrow flux-tube results, solid lines show medium flux-tube results, dotted lines show wide flux-tube results. (In the $^3S_1$ model, there is negligible dependence on the type of flux tube, so only a solid line is shown.) The arrows indicate our estimates of the actual FSI strengths. The shaded bands give the experimental ratios.

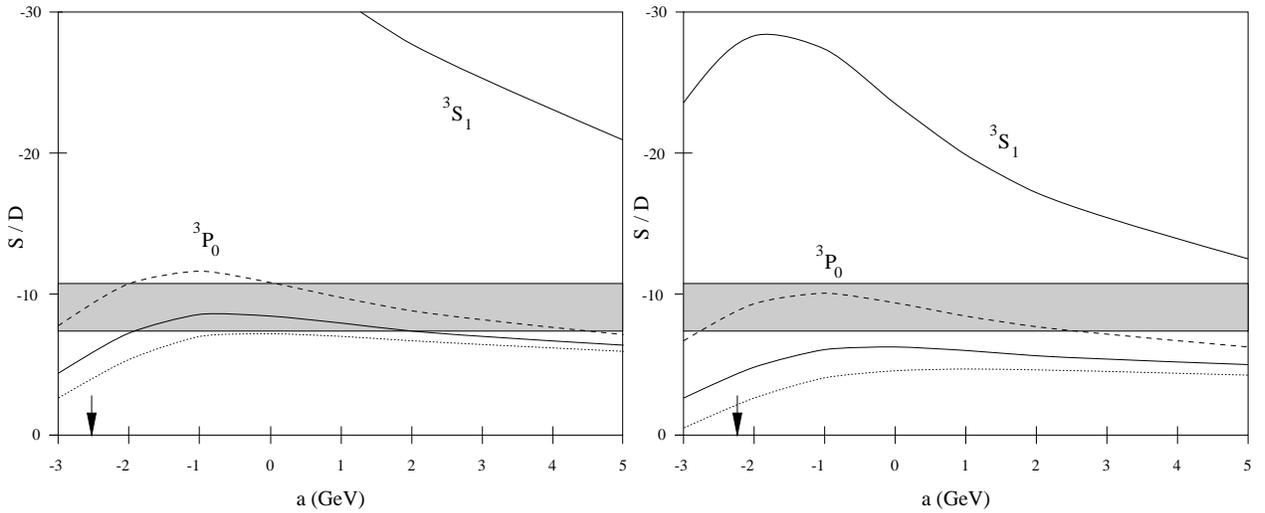

Fig. 3. $S/D$ amplitude ratios for $a_1 \to \pi\rho$. See Fig. 2 for key.

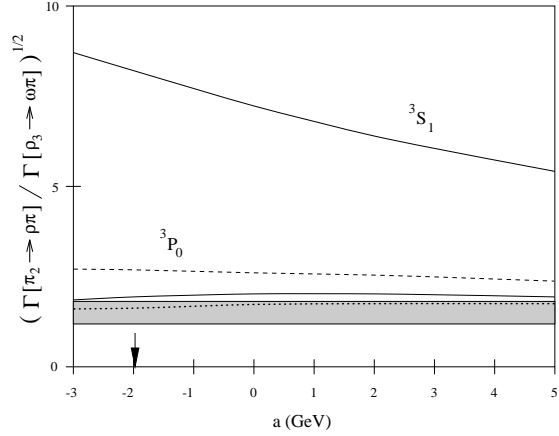

Fig. 4. The $(\pi_2 \to \rho\pi)/(\rho_3 \to \omega\pi)$ amplitude ratio (calculated with Coulomb-plus-linear wavefunctions). See Fig. 2 for key.

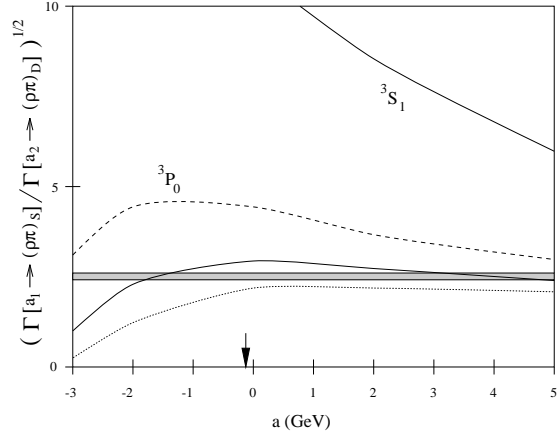

Fig. 5. The $(a_1 \to (\rho\pi)_S)/(a_2 \to \rho\pi)$ amplitude ratio (calculated with Coulomb-plus-linear wavefunctions). See Fig. 2 for key.

The situation for $\Gamma(a_1 \to (\rho\pi)_S)/\Gamma(a_2 \to \rho\pi)$ shown in Fig. 5 is similar, although here the various phase space factors differ by $\sim 10\%$. (The results shown were obtained using the KI phase space prescription.)

The $^3P_0$ model is strongly favoured by these results. However, ambiguities remain in its application to individual decay amplitudes. In particular the form of the phase space to be used is unclear. This is especially true for decay involving pions, where strong binding and relativistic effects can be very large. Due to its chiral nature, the pion is very light with respect to other hadrons, yet it interacts with typical hadronic strength. This combination of attributes is difficult to obtain in nonrelativistic models. For example the pion may be made as light as one wishes by increasing the strength of the hyperfine interaction, but only at the expense of decreasing the pion radius (and it is the size of hadrons which controls their interactions at low energy). Indeed, it seems that one reason for the success of the simple SHO wavefunctions is that they make the pion as large as the rho. Even so, special treatment must be given to decays involving pions in order to obtain agreement with measured widths. The Kokoski and Isgur phase space convention discussed in Section II provides one (very successful) method for enhancing pionic decay amplitudes, however it is of interest to check whether final state interactions can also provide the required enhancements. To this end, we have calculated the couplings necessary to reproduce experimental widths for a range of situations. The results are presented in Table II [24]. Case A corresponds to the 'standard' $^3P_0$ model, with SHO wavefunctions, no final state interactions, and no flux tube. It is apparent that the KI prescription gives the best results here. For comparison, Case B shows the couplings in the $^3S_1$ model with no final state interactions, SHO wavefunctions, and no flux tube. Here the required couplings are quite uniform except for $\gamma_S(\rho \to \pi\pi)$. Without FSI's, there is no single value of $\gamma_S$ which correctly reproduces all the widths.

The final two cases illustrate that it is possible to fit the decays using the $^3P_0$ model with relativistic (case C) or nonrelativistic (case D) phase space. Case C is corresponds to Coulomb plus linear wavefunctions with a narrow flux tube. The final state interactions have had their ranges increased by a factor of $\approx \sqrt{2}$ for each pion in the final state (Weinstein and Isgur have used a procedure similar to this in their study of the $\pi\pi$ and $K\bar{K}$ systems [25]; the idea is to compensate for the small pion radius. For an alternative viewpoint see Ref. [11]). In case D, a somewhat stronger $\pi\pi$ interaction (with a depth of 3.0 GeV; everything else as in case C) allows the nonrelativistic phase space to work well. Finally, it should be pointed out that the good fit to the amplitude ratios was maintained in all ($^3P_0$) cases.

Thus, we are unable to pin down a best set of parameters for the $^3P_0$ model; there are several combinations of phase space, flux tube width, and FSI strength which fit the data with comparable accuracy.

TABLE II. Coupling constants ($\gamma_P$ or $\gamma_S$) required to fit various decay widths.

| phase space | $\rho \to \pi\pi$ | $b_1 \to \pi\omega$ | $a_1 \to \pi\rho$ | $\pi_2 \to \pi\rho$ | $\bar{\gamma}$ | $\frac{\delta\gamma}{\bar{\gamma}}$ |
|---|---|---|---|---|---|---|
| Case A: $^3P_0$ model, SHO, No FSI's, No flux tube | | | | | | |
| $\tilde{m}_B \tilde{m}_C / \tilde{m}_A$ | 0.37 | 0.42 | 0.37 | 0.37 | 0.38 | 0.07 |
| $m_B m_C / m_A$ | 1.94 | 0.91 | 0.79 | 0.81 | 1.1 | 0.50 |
| $E_B E_C / m_A$ | 0.71 | 0.53 | 0.46 | 0.42 | 0.53 | 0.24 |
| Case B: $^3S_1$ model, SHO, No FSI's, No flux tube[a] | | | | | | |
| $\tilde{m}_B \tilde{m}_C / \tilde{m}_A$ | 0.89 | 0.23 | 0.19 | 0.23 | 0.39 | 0.86 |
| $m_B m_C / m_A$ | 4.8 | 0.49 | 0.42 | 0.51 | 1.56 | 1.39 |
| $E_B E_C / m_A$ | 1.7 | 0.28 | 0.25 | 0.26 | 0.62 | 1.16 |
| Case C: $^3P_0$ model, C+L, wide FSI's, Narrow flux tube | | | | | | |
| $\tilde{m}_B \tilde{m}_C / \tilde{m}_A$ | 2.8 | 4.4 | 4.8 | 4.5 | 4.1 | 0.22 |
| $m_B m_C / m_A$ | 15.4 | 9.5 | 10.4 | 10.0 | 11.3 | 0.24 |
| $E_B E_C / m_A$ | 5.5 | 5.5 | 6.0 | 5.1 | 5.5 | 0.07 |
| Case D: $^3P_0$ model, C+L, deep FSI's, Narrow flux tube | | | | | | |
| $\tilde{m}_B \tilde{m}_C / \tilde{m}_A$ | 2.0 | 4.4 | 4.8 | 4.5 | 3.9 | 0.33 |
| $m_B m_C / m_A$ | 10.9 | 9.5 | 10.4 | 10.0 | 10.2 | 0.06 |
| $E_B E_C / m_A$ | 3.9 | 5.5 | 6.0 | 5.1 | 4.9 | 0.24 |

[a] $\gamma_S$ has units of GeV$^{-1}$.

## IV. SUMMARY AND CONCLUSIONS

We have studied several meson decay amplitude ratios in the $^3S_1$ and $^3P_0$ models, using the Quark Born Formalism to calculate final state interactions between the decay products. These ratios, in which relativistic uncertainties cancel out almost completely, show a clear preference for the $^3P_0$ model; the $^3S_1$ model cannot reproduce the experimental data without unreasonably strong FSI's. A measurement of the $P/F$ amplitude ratio in $\pi_2 \to \pi\rho$ would further test this conclusion; we predict $P/F = 1.6 \pm .2$ using the $^3P_0$ model.

In addition, we have examined the parameter dependence of some individual decay amplitudes in the $^3P_0$ model. We found that there is no unique prescription for dealing with the above-mentioned relativistic ambiguities. That is, equally good results can be obtained using any of several prescriptions for the decay phase space.

Further progress will probably require a more fundamental description of the hadronic states and their couplings. A relativistic model which treats the decay and final state processes on an equal footing, and accommodates the special nature of the pion, would be extremely useful.

## ACKNOWLEDGMENTS

We are grateful to Nathan Isgur for encouragement. PG thanks NSERC of Canada and the U.S. Dept. of Energy under grant No. DE-FG02-91ER40682 for financial support. ES acknowledges support under DOE grant DE-AC02-76ER03069.

---

[1] Throughout this paper, we are concerned only with strong decays that are allowed by the Okubo-Zweig-Iizuka rule.

[2] A. Le Yaouanc, L. Oliver, O. Pene, and J.-C. Raynal, Phys. Rev. D **8**, 2233 (1973); Phys. Lett. **71 B**, 397 (1977); *ibid* **72 B**, 57 (1977). See also W. Roberts and B. Silvestre-Brac, Few Body Syst. **11**, 171 (1992).

[3] L. Micu, Nucl. Phys. **B10**, 521 (1969); R. Carlitz and M. Kislinger, Phys. Rev. D **2**, 336 (1970).

[4] R. Kokoski and N. Isgur, Phys. Rev. D **35**, 907 (1987).

[5] Fl. Stancu and P. Stassart, Phys. Rev. D **38**, 233 (1988); **39**, 343 (1989); **41**, 916 (1990); **42**, 1521 (1990).

[6] S. Capstick and W. Roberts, Phys. Rev. D **47**, 1994; CEBAF-TH-93-18 (unpublished).

[7] J.W. Alcock, M.J. Burfitt, W.N. Cottingham, Z. Phys. **C25**, 161 (1984).

[8] S. Kumano, Phys. Rev. D **41**, 195 (1990).

[9] M. Khono and W. Weise, Nucl. Phys. **A479**, 433c (1988).

[10] S. Kumano and V.R. Pandharipande, Phys. Rev. D **38**, 146 (1988).

[11] T. Barnes and E.S. Swanson, Phys. Rev. D **46**, 131 (1992)

[12] E.S. Swanson, Ann. Phys. (NY) **220**, 73 (1992).

[13] T. Barnes, E.S. Swanson, and J. Weinstein, Phys. Rev. D **46** (1992), 4868.

[14] T. Barnes and E.S. Swanson, to appear, Phys. Rev. C.

[15] T. Barnes, S. Capstick, M.D. Kovarik, and E.S. Swanson, Phys. Rev. C **48**, 539 (1993).

[16] H. Albrecht *et al.*, Z. Phys. C **58**, 61 (1993).

[17] N. Isgur and J. Paton, Phys. Rev. D **31**, 2910 (1985).

[18] Z. Li, M. Guidry, T. Barnes, and E.S. Swanson, *$I = 0, 1\ \pi\pi$ and $I = 1/2\ K\pi$ Scattering using Quark Born Diagrams*, MIT-CTP-2277/ORNL-CCIP-94-01, Jan. 1994.

[19] N.Isgur and R. Kokoski, Phys. Rev. Lett. **54**, 869 (1985); J. Merlin and J. Paton, J. Phys. G **11**, 439 (1985).

[20] For FSI calculations with SHO wavefunctions, we employ the delta function form of the hyperfine interaction (as opposed to the smeared delta function used in the Coulomb-plus-linear case). The confinement potential was of the form $-\frac{1}{2}\kappa \sum_{ij} r_{ij}^2 \frac{\vec{\lambda}_i}{2} \frac{\vec{\lambda}_j}{2}$. These choices keep the SHO results analytical and provide a further test of the stability of the effective potentials.

[21] It should be stressed that all of these potentials are valid for low momenta only; more complicated parameterizations may be necessary at higher momenta.

[22] *Review of Particle Properties*, Particle Data Group, Phys. Rev. D **45**, no. 11, (1992).

[23] The $S/D$ ratio for the $a_1$ is taken from the Argus measurement [16]; for consistency we also used the Argus $a_1$ mass of 1.211 MeV in our calculations.

[24] The measured widths are $\Gamma(\rho \to \pi\pi) = 151.5$ MeV, $\Gamma(b_1 \to \pi\omega) = 155$ MeV, $\Gamma(a_1 \to \pi\rho) = 446$ MeV, and $\Gamma(\pi_2 \to \pi\rho) = 77.5$ MeV.

[25] J. Weinstein and N. Isgur, Phys. Rev. D **41**, 2236 (1990).